\documentclass[a4paper]{jpconf}
\usepackage[latin9]{inputenc}
\usepackage{color}
\usepackage{amsmath}
\usepackage{amssymb}
\usepackage{graphicx}
\usepackage{esint}
\usepackage{rotating}
\usepackage{braket}
\usepackage{bbm}
\newcommand{\ens}[0]{\ensuremath}
\newcommand{\iE}[0]{\ens{\mathrm{i}}}

\makeatother
\begin{document}

\title{Cooperative quantum electrodynamical processes in an ellipsoidal
cavity}

\author{G. Alber, N. Trautmann}
\address{Institut für Angewandte Physik, Technische Universität Darmstadt,D-64289,
Germany}
\ead{nils.trautmann@physik.tu-darmstadt.de}

\date{\today}
\begin{abstract}
We investigate spontaneous photon emission and absorption processes
of two two-level atoms trapped close to the focal points of an ellipsoidal
cavity, thereby taking into account the full multimode scenario. In
particular, we calculate the excitation probabilities of the atoms
by describing the field modes semiclassically. Based on this approach,
we express the excitation probabilities by a semiclassical photon
path representation. Due to the special property of an ellipsoidal
cavity of having two focal points, we are able to study interesting
intermediate instances between well-known quantum-optical scenarios.
Furthermore, the semiclassical photon path representation
enables us to address the corresponding retardation effects and causality
questions in a straightforward manner. 
\end{abstract}

%\keywords{Quantum optics, Quantum information,  Quantum description of interaction of light and matter}
%\maketitle

\section{Introduction}

During the last decades, the field of quantum optics has witnessed
remarkable experimental developments. They have enabled new possibilities
of studying resonant light-matter interaction \cite{Berman1994,Walther2006,Haroche2006}.
These developments are not only interesting from a fundamental point
of view, but also from an applied perspective, because they are relevant
for advanced quantum technologies and their possible application in
quantum information processing.

Understanding light-matter interaction is crucial for being able to
transfer quantum information between single photons, frequently used
as flying qubits, and elementary material systems, typically used
as stationary qubits in applications in quantum information. Such
a transfer of quantum information between single photons and elementary
material systems, such as two level systems serving as qubits, is
an integral building block for achieving quantum communication over
large distances. Recently, considerable experimental effort \cite{Goy1983,Meschede1985,McKeever2003}
has been devoted to investigating the interaction of matter qubits
with one or a few selected modes of the radiation field within the
framework of the Jaynes-Cummings-Paul model \cite{Schleich,jaynes1963comparison}.
Recent experimental work has extended these quantum electrodynamical
scenarios to the opposite limit of extreme multimode scenarios \cite{Moehring2007,Olmschenk2009,PhysRevLett.111.103001}
with structured continua of electromagnetic field modes characteristic
for half open cavities, such as a parabolic mirror\cite{Maiwald2009,Maiwald2012,hetet2011single}.

Motivated by these latter developments, we explore in this paper basic
dynamical features of quantum electrodynamic processes in extreme
multimode scenarios. In particular, we investigate the almost resonant
interaction between two two-level systems located at the focal points
of a prolate-ellipsoidal cavity and the quantized electromagnetic
field. A particularly interesting elementary quantum electrodynamical
situation arises if these two atoms exchange a single photon emitted
spontaneously by one of these two-level systems. Due to the special
property of the prolate-ellipsoidal cavity of having two focal points,
the electromagnetic field strength around the focal points is enhanced
significantly, thus causing an interesting interaction between the
two-level systems and the radiation field. From the point
of view of the two-level systems as an open quantum system, the resulting
reduced dynamics is highly non-Markovian with significant memory effects.
Thus, an ellipsoidal cavity is an interesting scenario to study almost
resonant matter field interaction. In particular, by changing the
size of the cavity continuously, all intermediate cases between the
single mode scenario, as described by the Jaynes-Cummings-Paul model,
the large cavity limit, and the case of a structured continua of field
modes can be addressed. Even the recently explored case of a parabolic
cavity \cite{Alber2013} can be reproduced in the limit of infinitely
separated focal points. Furthermore, we are able to tune the coupling
strength between the two-level systems and the radiation field by
changing the shape of the cavity. In addition, we are able to investigate
diffraction effects in the regime of wavelengths large in comparison
to the characteristic length scales inside the cavity and their extinction
in the geometric optical limit of small wavelengths. In the following
it is demonstrated that all these effects can be described adequately
with the help of a semiclassical description of the mode structure
inside the ellipsoidal cavity. We show that these semiclassical methods
lead to a convenient photon path representation by which all relevant
quantum mechanical transition amplitudes are expressed as a linear
superposition of contributions of relevant photon paths inside the
cavity. 

This paper is organized as follows. In Sec. \ref{quantum electrodynamical model}
we introduce our theoretical model and the main approximations involved.
The Helmholtz equation with the appropriate boundary conditions is
solved by semiclassical methods in Sec. \ref{Solving the Helmholtz equation by using semiclassical methods}.
The photon path representation for describing the time evolution of
relevant quantum mechanical transition amplitudes is presented in
Sec. \ref{Photon_path_representation}. Numerical results for the
dynamics of the two-level systems are discussed in Sec. \ref{Results}.

\section{Quantum electrodynamical model\label{quantum electrodynamical model}}

\begin{figure}[t]
\begin{centering}
\includegraphics[scale=0.15]{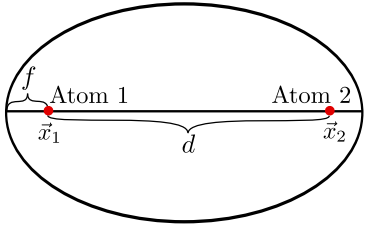}
\par\end{centering}

\caption{Two two-level systems located in the foci of a prolate-ellipsoidal
cavity. \label{fig:Setup}}
\end{figure}

We investigate the dynamics of two identical two-level systems, e.g.
atoms or ions, situated at the focal points $\mathbf{x}_{a}$ ($a\in\left\{ 1,2\right\} $)
of an ideally conducting prolate-ellipsoidal cavity as shown schematically
in Fig. \ref{fig:Setup}. Both atoms interact with the quantized radiation
field inside this cavity. The two two-level systems are assumed to
be trapped in such a way that their center of mass motion is negligible.
We assume that the dipole matrix elements $\mathbf{d}_{a}=\bra{e}_{a}\hat{\mathbf{d}}_{a}\ket{g}_{a}$
are oriented along the symmetry axis, i.e. $z-$axis, of the system
so that we may write 
\begin{equation}
\mathbf{d}_{a}=D\mathbf{e}_{z}\;\text{with }D\in\mathbb{R}.
\end{equation}

The identical transition frequencies $\omega_{eg}$ of the two-level
systems are assumed to be in the optical frequency regime and to dominate
the coupling to the radiation field as measured by relevant Rabi frequencies,
for example. Furthermore, the sizes of the electronic states involved
in the transitions of the two-level systems are supposed to be small
compared to the corresponding wavelength $\lambda_{\text{eg}}=2\pi c_{0}/\omega_{eg}$
of the transition. ($c_{0}$ being the speed of light in vacuum.)
Thus, the dipole and the rotating-wave approximation (RWA) are applicable.
Therefore, the quantum electrodynamical interaction between the two
two-level systems and the quantized electromagnetic field inside the
cavity is described by the Hamiltonian 
\begin{equation}
\hat{H}=\hat{H}_{\text{atoms}}+\hat{H}_{\text{field}}+\hat{H}_{\text{i}}\label{eq:Hamiltonian}
\end{equation}
with $\hat{H}_{\text{field}}=\sum_{i}\hbar\omega_{i}\hat{a}_{i}^{\dagger}\hat{a}_{i}$,
$\hat{H}_{\text{atoms}}=\hbar\omega_{eg}\sum\limits _{a=1}^{2}\ket{e}_{a}\bra{e}_{a},$
$\hat{H}_{\text{i}}=-\sum\limits _{a=1}^{2}\hat{\mathbf{E}}_{\perp}^{+}(\mathbf{x}_{a})\cdot\hat{\mathbf{d}}_{a}^{-}+\text{H.c.}$,
and with the dipole transition operator $\hat{\mathbf{d}}_{a}^{-}=\mathbf{d}_{a}^{*}\ket{g}_{a}\bra{e}_{a}$
of the two-level atom $a\in\left\{ 1,2\right\} $. In the Schrödinger
picture the positive frequency part of the (transversal) electromagnetic
field operator is given by 
\begin{equation}
\hat{\mathbf{E}}_{\perp}^{+}(\mathbf{x})=-\iE\sum_{i}\sqrt{\frac{\hbar\omega_{i}}{2\epsilon_{0}}}\mathbf{g}_{i}(\mathbf{x})\hat{a}_{i}^{\dagger}
\end{equation}
with $\mathbf{g}_{i}(\mathbf{x})$ denoting the orthonormal transversal
electric mode functions, i.e. $\int_{V}d^{3}{\bf x}~\mathbf{g}_{i}^{*}(\mathbf{x})\cdot\mathbf{g}_{j}(\mathbf{x})=\delta_{ij}$,
which are assumed to fulfill the boundary conditions for an ideally
conducting cavity.

\section{Mode functions of the cavity and their semiclassical approximation
\label{Solving the Helmholtz equation by using semiclassical methods}}

\subsection{Separation ansatz for the relevant mode functions\label{Seperatio_ansatz}}

In order to determine the electric mode functions by solving the Helmholtz
equation, we are going to use prolate-ellipsoidal coordinates. The
connection between the prolate-ellipsoidal coordinates and the Cartesian
coordinates is given by 
\begin{equation}
\left(\begin{array}{c}
x\\
y\\
z
\end{array}\right)=\frac{d}{2}\left(\begin{array}{c}
\cos(\varphi)\sqrt{(1-\eta^{2})(\xi^{2}-1)}\\
\sin(\varphi)\sqrt{(1-\eta^{2})(\xi^{2}-1)}\\
\eta\xi
\end{array}\right).
\end{equation}
Thereby, the focal points have been chosen to be located at the positions
$\pm\mathbf{e}_{z}d/2$ and $d$ denotes their distance. The ranges
of the prolate-ellipsoidal coordinates are given by $\varphi\in[0,2\pi)$,
$\eta\in[-1,1]$, and $\xi\in[1,\infty)$. In particular, the focal
points have the coordinates $\xi=1$, $\eta=\pm1$. In these new coordinates
an ellipsoid with the focal length $f$ is defined by the surface
\begin{equation}
\xi=2f/d+1\;.
\end{equation}

Only those mode functions enter the interaction Hamiltonian $\hat{H}_{\text{i}}$
which have a nonvanishing scalar product with the dipole transition
operators of the two two-level systems situated in the focal points
of the cavity. By taking into account the transversality of the radiation
field, it turns out that for dipole matrix elements oriented along the $z$-axis
all these modes can be obtained by a separation ansatz.
This separation ansatz is given by
\begin{equation}
\mathbf{g}_{i}(\varphi,\xi,\eta)=\nabla\times H,\; H=\frac{1}{\mathcal{N}}\mathbf{e}_{\varphi}\mathcal{F}_{i}(\kappa,\xi)\mathcal{G}_{i}(\kappa,\eta)\;,
\end{equation}
with \begin{footnotesize}
\begin{eqnarray}
\left[\alpha_{n}\kappa-\kappa^{2}\left(\eta^{2}-1\right)-\frac{1}{1-\eta^{2}}\right]\mathcal{G}(\kappa,\eta)+\frac{d}{d\eta}(1-\eta^{2})\frac{d}{d\eta}\mathcal{G}(\kappa,\eta)&=&0\;,\nonumber\\\label{eq:DGL_S_eta}\\
\left[\alpha_{n}\kappa-\kappa^{2}\left(\xi^{2}-1\right)-\frac{1}{1-\xi^{2}}\right]\mathcal{F}(\kappa,\xi)+\frac{d}{d\xi}(1-\xi^{2})\frac{d}{d\xi}\mathcal{F}(\kappa,\xi)&=&0\;,\nonumber\\\label{eq:DGL_R_xi}
\end{eqnarray}
\end{footnotesize}
whereby $\kappa=\omega d/(2c_{0})$ , $\alpha_{n}$
being the constant of separation and $\mathcal{N}$ being the normalization
factor which has to be chosen such that
\begin{equation}
\int\limits _{V}\mathbf{g}_{i}(x)\cdot\mathbf{g}_{j}^{*}(x)d^{3}x=\delta_{i,j}\;\forall i,j\;.\label{eq:Normalizatio_condition}
\end{equation}
The index $n$ indicates that $\mathcal{G}(\kappa,\eta)$ remains finite
for $\eta\rightarrow\pm1$ only for countable sets of the separation constant $\alpha_{n}$.
The differential equations Eq. (\ref{eq:DGL_S_eta}) and Eq. (\ref{eq:DGL_R_xi})
are actually identical, and the functions $\mathcal{G}$ and
$\mathcal{F}$ just differ by the domains for $\eta$ and $\xi$.
The solutions of these differential equations are prolate-spheroidal
wave functions \cite{Flammer1957}. 

\subsection{Uniform semiclassical approximation\label{Uniform_JWKB}}

In the previous section, we have expressed the mode functions of the
electric field operator in terms of prolate-spheroidal wave functions.
We can obtain an analytical solution for this equations in the special case
of $\alpha_{n}=0$ 
\begin{eqnarray}
\mathcal{G}(\kappa,\eta)\Bigr\vert_{\alpha_{n}=0} & = & \frac{\text{sin}[\kappa(1-\eta)]}{\sqrt{1-\eta^{2}}},\label{eq:G_exact}\\
\mathcal{F}(\kappa,\xi)\Bigr\vert_{\alpha_{n}=0} & = & \frac{\text{sin}[\kappa(\xi-1)]}{\sqrt{\xi^{2}-1}}\;.\label{eq:F_exact}
\end{eqnarray}
A general closed analytical expression for arbitrary values
is not known. However, with the help of semiclassical methods, approximate
expressions can be found. It turns
out that the mode functions with $\alpha_{n}\approx0$ yield the dominant contributions.
Therefore, we can exploit the exact analytical
solutions for $\alpha_{n}=0$ in order to improve our approximation.
It is convenient to choose the following
normalization conditions for the functions $\mathcal{F}(\kappa,\xi)$ and
$\mathcal{G}(\kappa,\eta)$ 
\begin{eqnarray}
\mathcal{F}(\kappa,\xi) & \underset{\kappa\xi\rightarrow\infty}{\longrightarrow} & \frac{1}{\xi}\cos\left[\kappa\xi-\frac{1}{2}\left(n+1\right)\pi\right],\nonumber \\
\mathcal{G}(\kappa,0) & = & (-1)^{(n+1)/2}\;\text{in case of odd }n,\nonumber \\
\frac{d}{d\eta}\mathcal{G}(\kappa,\eta)\Bigr\vert_{\eta=0} & =&(-1)^{n/2}  \kappa\;\text{in case of even }n.\nonumber\\\label{normalization}
\end{eqnarray}
These conditions are fulfilled by the exact solutions in Eq. (\ref{eq:G_exact})
and Eq. (\ref{eq:F_exact}).

In order to circumvent problems originating from the singularities
of the differential equations (\ref{eq:DGL_S_eta}) and (\ref{eq:DGL_R_xi}),
we first of all apply a transformation which removes these singularities
at the positions of the focal points. Let us demonstrate this procedure
by concentrating on the functions $\mathcal{F}(\kappa,\xi)$. The
procedure for the functions $\mathcal{G}(\kappa,\eta)$ is analogous
because the differential equations for $\mathcal{F}$ and for $\mathcal{G}$
coincide. A transformation removing the singularities is given by
\begin{equation}
\mathcal{F}(\kappa,\xi)=\frac{1}{\sqrt{\xi^{2}-1}}f(\sqrt{\xi^{2}-1})\;.
\end{equation}
Consequently, the function $f(x)$ with $x\in[0,\infty)$ is a solution
of the differential equation 
\begin{equation}
0=\kappa^{2}x\left(x^{2}-\frac{\alpha_{n}}{\kappa}\right)f(x)-f'(x)+x\left(1+x^{2}\right)f''(x)\label{eq:Differentialgleichung_exakt}
\end{equation}
with $x=\sqrt{\xi^{2}-1}$. The focal points correspond to $x=0$.
For the determination of the relevant mode functions, we are only interested
in the regular (physical) solution of Eq. (\ref{eq:Differentialgleichung_exakt})
which remains finite for all possible values of $x\geq0$.

A uniform semiclassical JWKB-approximation \cite{Berry1972} for the
solution of Eq. (\ref{eq:Differentialgleichung_exakt}) can be constructed
by using the differential equation 
\begin{equation}
\kappa^{2}\sigma\left(\sigma^{2}-\frac{\alpha_{n}}{\kappa}\right)\tilde{f}(\sigma)-\tilde{f}'(\sigma)+\sigma\tilde{f}''(\sigma)=0\;,\label{eq:Differentialgleichung_approx}
\end{equation}
as a comparison equation with the exact regular solution 
\begin{equation}
\tilde{f}(\sigma)=e^{-\iE\kappa\sigma^{2}/2}\kappa  \sigma ^2 \, _1F_1\left(1-\iE\alpha_{n}/4;2;i \kappa  \sigma ^2\right)\;.\label{eq:Comparsion_solution}
\end{equation}
In Eq. (\ref{eq:Comparsion_solution}) ${}_1F_1$ denotes Kummer's confluent hypergeometric function \cite{AS}.
Accordingly,  we are going to solve Eq. (\ref{eq:Differentialgleichung_exakt})  by constructing a sufficiently smooth
mapping between the variables $x$ and $\sigma$ in such a way that
fulfillment of Eq. (\ref{eq:Differentialgleichung_exakt}) is approximately
equivalent to fulfillment of Eq. (\ref{eq:Differentialgleichung_approx}).
For this purpose, we transform Eq. (\ref{eq:Differentialgleichung_exakt})
into a canonical form by eliminating the terms involving first order
derivates. This is achieved by the transformation $f(x)=\phi(x)u(x)$
with 
\begin{equation}
\phi(x)=\frac{\sqrt{x}}{\left(1+x^{2}\right)^{1/4}}.
\end{equation}
The resulting canonical form of Eq. (\ref{eq:Differentialgleichung_exakt})
is given by 
\begin{eqnarray}
u''(x)+\chi(x)u(x) & = & 0\label{canonical}
\end{eqnarray}
with 
\begin{equation}
\chi(x)=\frac{2x^{2}\left(2\kappa\left(x^{2}+1\right)\left(\kappa x^{2}-\alpha_{n}\right)-3\right)-3}{4\left(x^{3}+x\right)^{2}}\;.
\end{equation}
In order to find a uniform JWKB-approximation for Eq. (\ref{canonical}),
we have to chose an appropriate comparison equation whose solution
is known exactly. For our purposes, we chose the comparison equation
\begin{eqnarray}
\tilde{u}''(\sigma)+\Pi(\sigma)\tilde{u}(\sigma) & = & 0\label{comparison}
\end{eqnarray}
with 
\begin{equation}
\Pi(\sigma)=\kappa^{2}\sigma^{2}-\alpha_{n}\kappa-\frac{3}{4\sigma^{2}}.
\end{equation}
This comparison equation is related to the differential equation (\ref{eq:Differentialgleichung_approx})
by the relation $\tilde{f}(\sigma)=\sqrt{\sigma}\tilde{u}(\sigma)$.
Furthermore, we have to choose a smooth mapping between the independent
variables $x$ and $\sigma$ determined by the relation 
\begin{eqnarray}
\frac{d\sigma}{dx} & = & \sqrt{\frac{\chi(x)}{\Pi(\sigma(x))}}\label{mapping}\;.
\end{eqnarray}
The initial condition for the solution of Eq. (\ref{mapping}) has to
be chosen such that the first positive zero of $\chi(x)$,
i.e. $x_{0}$, is mapped onto the first positive zero of $\Pi(\sigma)$,
i.e. $\sigma_{0}$, in order to avoid singularities. Solving Eq. (\ref{mapping}) analytically is a
complicated problem. However, simple
expressions are available for $x\rightarrow0$ and $x\rightarrow\infty$ which are given by $\sigma(x)\underset{x\rightarrow0}{\longrightarrow}x$
and $\sigma(x)\underset{x\rightarrow\infty}{\longrightarrow}\sqrt{2x}$.
The behavior for $x\rightarrow0$ is of importance in order to evaluate
the mode function near the focal points and $x\rightarrow\infty$
is relevant for the normalization (Eq. (\ref{normalization})). 
Thus, the
JWKB-approximation for $\mathcal{F}(\kappa,\xi)$ is finally given
by  
\begin{eqnarray}
&\mathcal{F}(\kappa,\xi)=\frac{1}{\left(\xi^{2}\left(-1+\xi^{2}\right)\right)^{1/4}}{\left(\frac{\Pi(\sigma(\sqrt{\xi^{2}-1}))}{\chi(\sqrt{\xi^{2}-1})}\right)}^{1/4}\psi(\sigma(\sqrt{\xi^{2}-1}))\nonumber\\
&
\end{eqnarray}
with \cite{AS}
\begin{equation}
\psi(\sigma)=e^{-\pi\alpha_{n}/8}\text{ }\sqrt{\frac{\pi\alpha_{n}}{16\sigma}\text{csch}\left(\frac{\pi\alpha_{n}}{4}\right)}\tilde{f}(\sigma).
\end{equation}
This expression for $\mathcal{F}(\kappa,\xi)$ fulfills the normalization condition Eq. (\ref{normalization}). The normalized mode function at the first focal point is now given by
\begin{eqnarray}
&\mathbf{g}_{i}(\varphi,\xi,\eta)\Bigr\vert_{\eta=1,\xi=1}=\nonumber\\
&\mathbf{e}_{z}\frac{4}{d\mathcal{N}(\kappa,\alpha_{n})}\left(\lim\limits _{\xi\rightarrow1}\frac{\mathcal{F}(\kappa,\xi)}{\sqrt{\xi^{2}-1}}\right)
\left(\lim\limits _{\eta\rightarrow1}\frac{\mathcal{G}(\kappa,\eta)}{\sqrt{1-\eta^{2}}}\right)\nonumber\\
&=\mathbf{e}_{z}\frac{\kappa^{2}\pi}{4d\mathcal{N}(\kappa,\alpha_{n})}\alpha_{n}\text{csch}\left({\pi\alpha_{n}}/{4}\right)\;.\label{coupling}
\end{eqnarray}

By exploiting the symmetry of the problem, we can evaluate the mode function at the second
focal point and obtain

\begin{equation}
\mathbf{g}_{i}(\varphi,\xi,\eta)\Bigr\vert_{\eta=-1,\xi=1}=(-1)^{n}\mathbf{g}_{i}(\varphi,\xi,\eta)\Bigr\vert_{\eta=1,\xi=1}\;.\label{coupling1}
\end{equation}
In order to normalize the mode functions according to Eq. (\ref{eq:Normalizatio_condition})
we have to determine the constant $\mathcal{N}(\kappa,\alpha_{n})$, i.e.
\begin{equation}
\mathcal{N}(\kappa,\alpha_{n})=\kappa\sqrt{d\pi(I_{\mathcal{F}}^{2}I_{\mathcal{G}}^{0}-I_{\mathcal{F}}^{0}I_{\mathcal{G}}^{2})}
\end{equation}
with ($p=0,2$)
\begin{eqnarray}
I_{\mathcal{G}}^{p}=2\int\limits _{0}^{1}\mathcal{G}(\kappa,\eta)^{2}\eta^{p}d\eta\text{ and }I_{\mathcal{F}}^{p}=\int\limits _{1}^{1+2f/d}\mathcal{F}(\kappa,\xi)^{2}\xi^{p}d\xi.\nonumber\\    
\end{eqnarray}
According to Eqs. (\ref{coupling}) and (\ref{coupling1})  at the focal points the normalized mode functions are determined by the function $\mathcal{N}(\kappa,\alpha_{n})$ and the function $\alpha_{n}\text{csch}\left(\pi\alpha_{n}/4\right)$.
The function $\alpha_{n}\text{csch}\left(\pi\alpha_{n}/4\right)$ significantly
deviates from 0 only in the region around $\alpha_{n}=0$. It turns out that in this region $\mathcal{N}(\kappa,\alpha_{n})$ is slowly varying in comparison with the function $\alpha_{n}\text{csch}\left(\pi\alpha_{n}/4\right)$ in case of $\lambda_{eg}\ll d,f$.
This can be verified by using
the semiclassical potential $\chi(x)$ and a corresponding
potential for the semiclassical treatment of $\mathcal{G}$.
Thus the modes  with $\alpha_n\approx 0$ are the ones of main importance as far as their coupling to the dipoles of the
two-level systems is concerned and
$\mathcal{N}(\kappa,\alpha_{n})$
can be assumed to be independent of $\alpha_{n}$.
In addition it is also well-known that the relevant modes have the
property $\omega\approx\omega_{eg}$ which directly translates to
$\kappa\approx\kappa_{eg}=2\omega_{eg}d/c_{0}$ . By incorporating
these facts we can approximate
$\mathcal{N}(\kappa,\alpha_{n})$ by $\mathcal{N}(\kappa_{eg},0)$ in the subsequent calculation.

\subsection{Semiclassical quantization functions \label{quantization_functions}}

In the previous subsection, we have identified the modes in the region
of $\alpha_{n}\approx0$, $\kappa\approx\kappa_{eg}$ as the ones
of main importance for the dynamics of the system. However, in the case of $\alpha_{n}=0$
, Eqs. (\ref{eq:DGL_S_eta}) and (\ref{eq:DGL_R_xi}) can be solved
analytically. This particular feature of the mode functions can be exploited
for incorporating the boundary conditions of an ideally conducting
metallic boundary and determining the corresponding quantization functions
semiclassically. These quantization functions $n_{1}(\alpha_{n},\kappa),$
and $n_{2}(\alpha_{n},\kappa)$ are defined in such a way that the condition
\begin{equation}
\left(n_{1}(\alpha_{n},\kappa),n_{2}(\alpha_{n},\kappa)\right)\in\mathbb{N}_{0}\times\mathbb{N}_{0} 
\end{equation}
determines the possible mode functions which fulfill the boundary
conditions.

The quantization function $n_{1}(\alpha_{n},\kappa)$
is determined by imposing the condition that the function $\mathcal{G}(\kappa,\eta)$
has to remain finite for  $\eta\rightarrow\pm1$. The
quantization function $n_{2}(\alpha_{n},\kappa)$ takes into account
the boundary conditions of an ideally conducting surface of the cavity.
In particular, this implies that the tangential components of the
electric field strength and the normal components of the magnetic
field strength have to vanish at the boundary
which yield the constraint
\begin{equation}
 0=\frac{\partial}{\partial\xi}\left[\sqrt{\xi^{2}-1}\mathcal{F}(\kappa,\xi)\right]{}_{\xi={2f}/{d}+1}\;.
\end{equation}
As mentioned above we have to evaluate the quantization functions
around $\alpha_{n}\approx0$ and $\kappa\approx\kappa_{eg}$.
Therefore,
we can approximate the quantization functions $n_{1}(\alpha_{n},\kappa),\; n_{2}(\alpha_{n},\kappa)$
by their first order Taylor expansions in $\alpha_{n}$ and $\kappa$
around the values $\alpha_{n}=0$ and $\kappa=\kappa_{eg}$.
% \begin{eqnarray}
% &n_{i}(\alpha_{n},\kappa) \approx\nonumber
% n_{i}(0,\kappa_{eg})+\frac{\partial n_{i}}{\partial\alpha_{n}}\Bigr\vert_{\begin{array}{c}
% \alpha_{n}=0,\\
% \kappa=\kappa_{eg}
% \end{array}}\alpha_{n}\nonumber\\
% &+\frac{\partial n_{i}}{\partial\kappa}\Bigr\vert_{\begin{array}{c}
% \alpha_{n}=0,\\
% \kappa=\kappa_{eg}
% \end{array}}\left(\kappa-\kappa_{eg}\right)\;.
% \end{eqnarray}
This Taylor expansion is given by
\begin{eqnarray}
&n_{i}(\alpha_{n},\kappa) \approx n_{i}(0,\kappa_{eg})+\partial_{\alpha_{n}}n_{i}\alpha_n+\partial_{\kappa}n_{i}\left(\kappa-\kappa_{eg}\right)\nonumber\\
\end{eqnarray}
with $\partial_{\alpha_{n}}n_{i}$ and $\partial_{\kappa}n_{i}$ denoting the partial derivatives of the function $n_{i}(\alpha_{n},\kappa)$ with respect to the variable $\alpha_{n}$ respectively $\kappa$ for $\alpha_{n}=0$ and $\kappa=\kappa_{eg}$.
By exploiting the exact analytical expressions for $\mathcal{G}(\kappa,\eta)$
and $\mathcal{F}(\kappa,\xi)$ for $\alpha_{n}=0$ we obtain the relations 
\begin{equation}
n_{1}(0,\kappa_{eg})+\partial_{\kappa}n_{1} \left(\kappa-\kappa_{eg}\right)=\frac{2\kappa}{\pi}
\end{equation}
and 
\begin{equation}
n_{2}(0,\kappa_{eg})+\partial_{\kappa}n_{2}\left(\kappa-\kappa_{eg}\right)
=\frac{2\kappa f}{\pi d}+\frac{1}{2}\;.
\end{equation}
In order to specify the quantization functions in the linearizion
approximation completely, we still have to determine $\partial_{\alpha_{n}}n_{i}$.
This is achieved by invoking the semiclassical quantization condition. In case of a simple JWKB-approximation with Langer substitution \cite{Berry1972} the quantization function and the function $\mathcal{G}(\eta)$
  are given by
\begin{eqnarray}
 n_{1}&=&2 S(0)/\pi-\frac{1}{2}\\
 \mathcal{G}(\eta)&=&\sqrt[4]{\frac{V(0)}{V(\eta)}}\frac{1}{\sqrt{1+\eta}}\sin( S(\eta)+\frac{\pi}{4})
\end{eqnarray}
with
\begin{eqnarray}
S(\eta)=\kappa\int\limits _{\eta}^{\eta_{\text{turn}}}\sqrt{-V(\eta)}\frac{1}{1-\eta}d\eta
\end{eqnarray}
denoting the semiclassical eikonal equation. The quantity
\begin{eqnarray}
V(\eta)=\frac{1}{4\kappa^{2}}-\frac{(\eta-1)(\eta^{2}-1-\alpha_{n}/\kappa)}{1+\eta}
\end{eqnarray}
is the semiclassical potential and $\eta_{\text{turn}}$ denotes the turning point of the semiclassical
potential $V(\eta)$ in the region $[0,1]$ ($V(\eta_{\text{turn}})=0$). Thereby, we obtain 

 \begin{eqnarray}
&\partial_{\alpha_{n}}n_{1}=\frac{1}{\pi}\int\limits _{0}^{\eta_{\text{turn}}}\frac{1}{\sqrt{-V(\eta)}}\frac{1}{1+\eta}d\eta\nonumber\nonumber\\
&\approx\frac{1}{\pi \sqrt{-V(0)}}\int\limits _{-1}^{1}\mathcal{G}(\eta)^{2}d\eta= I_{\mathcal{G}}^{0}/\left(\pi\sqrt{1-\frac{1}{4\kappa_{eg}^{2}}}\right)\;\nonumber\\
\end{eqnarray}
whereby all expressions have to be evaluated for $\alpha_{n}=0$ and $\kappa=\kappa_{eg}$.
We can apply the same procedure to $\partial_{\alpha_{n}}n_{2}$
and obtain 
\begin{equation}
 \partial_{\alpha_{n}}n_{2}\approx-\frac{1}{\pi}I_{\mathcal{F}}^{0}\;.
\end{equation}
We just have to evaluate $I_{\mathcal{F}}^{p}$ and $I_{\mathcal{G}}^{p}$
for $\alpha_{n}=0$ and $\kappa_{eg}$ which can
again be done by using Eq. (\ref{eq:G_exact}) and Eq. (\ref{eq:F_exact}), respectively.
From now on we will denote with $I_{\mathcal{F}}^{p}$ and $I_{\mathcal{G}}^{p}$
the values for $\alpha_{n}=0$ and $\kappa=\kappa_{eg}$.

\section{Photon path representation\label{Photon_path_representation}}

\subsection{Solving the Schrödinger equation by a photon path representation\label{Solving_Schroedinger}}

We can use the results of the previous section in order to determine
the time evolution of the system.
If we assume that initially only one two-level
atom is in an excited state and that the field is in the vacuum state,
the time evolution of the system is restricted to the subspace of
the Hilbert space which corresponds one excitation only.
Each state in this subspace is covered by the following
ansatz for the wave function 
\begin{eqnarray}
&\ket{\psi(t)}=b^{1}(t)\ket{e,g}^{A}\ket{0}^{P}+b^{2}(t)\ket{g,e}^{A}\ket{0}^{P}\nonumber\\
&+\sum\limits _{i}f_{i}(t)\ket{g,g}^{A}\hat{a}_{i}^{\dagger}\ket{0}^{P}\;.
\end{eqnarray}
(The superscripts $A$ and $P$ refer to atoms and photons, respectively.)
The Schrödinger equation leads to
a coupled system of linear differential equations. We apply the Laplace
transform in order to obtain a system of linear algebraic equations.
Hereby, we define the Laplace transforms of the probability amplitudes
by 
\begin{eqnarray}
&\ket{\widetilde{\psi}(\Lambda)}=\int\limits _{0}^{\infty}e^{\iE\Lambda t}\ket{\psi(t)}dt=\widetilde{b^{1}}(\Lambda)\ket{e,g}^{A}\ket{0}^{P}\nonumber\\
&+\widetilde{b^{2}}(\Lambda)\ket{g,e}^{A}\ket{0}^{P}+\sum\limits _{i}\widetilde{f_{i}}(\Lambda)\ket{g,g}^{A}\ket{1}_{i}^{P}
\end{eqnarray}
 for $\text{Im}(\Lambda)>0$. By eliminating the Laplace
transforms of the photonic excitations $\widetilde{f}_{i}(\Lambda)$,
we obtain
\begin{eqnarray}
&\iE\left(\begin{array}{c}
b^{1}(0)\\
b^{2}(0)
\end{array}\right)\nonumber\\
&=\left[\Lambda-\omega_{eg}+\iE\left(\begin{array}{cc}
A^{1,1}(\Lambda) & A^{1,2}(\Lambda)\\
A^{2,1}(\Lambda) & A^{2,2}(\Lambda)
\end{array}\right)\right]\left(\begin{array}{c}
\widetilde{b^{1}}(\Lambda)\\
\widetilde{b^{2}}(\Lambda)
\end{array}\right)\nonumber\\
\end{eqnarray}
with 
\begin{eqnarray}
&A^{a,b}(\Lambda)=\iE|D|^{2}\sum\limits _{j}\frac{\omega_{j}}{2\epsilon_{0}\hbar}\frac{\left(\mathbf{g}_{j}(\mathbf{x}_{a})\right)_{z}\left(\mathbf{g}_{j}(\mathbf{x}_{b})\right)_{z}^{*}}{\Lambda-\omega_{j}}\; a,b\in\left\{ 1,2\right\} \;.\nonumber\\
\end{eqnarray}
All properties of the cavity are now encoded in the functions
$A^{a,b}$. It is a non trivial task to evaluate these functions because
they are defined by an infinite sum over all modes $j$ which couple
to the atomic dipoles. However, by exploiting the semiclassical expressions
derived previously, we are able to evaluate these functions. This way,
we can solve the linear system of equations. Nevertheless,
evaluating the inverse Laplace transform is still a non trivial task.
For this purpose it is convenient to solve the linear system of equations
by applying the Neumann series. The corresponding solution is given
by
\begin{eqnarray}
&\left(\begin{array}{c}
\widetilde{b^{1}}(\Lambda)\\
\widetilde{b^{2}}(\Lambda)
\end{array}\right)
&=\iE\sum\limits_{n=0}^{\infty}\frac{\left[-\iE\left(\begin{array}{cc}
T_{1}^{1,1} & T_{1}^{1,2}\\
T_{1}^{2,1} & T_{1}^{2,2}
\end{array}\right)\right]^{n}}{(\Lambda+\iE{\Gamma}/{2}-\omega_{eg})^{n+1}}\left(\begin{array}{c}
b^{1}(0)\\
b^{2}(0)
\end{array}\right)\;.\nonumber\\\label{eq:Photon_path_representation}
\end{eqnarray}
Thereby, we have defined $T_{1}^{a,b}=A^{a,b}(\Lambda)-\delta_{a,b}\Gamma/2$
with $\Gamma$ denoting the spontaneous decay rate of the excited state $\ket{e}$.
In order to enforce convergence of the series in Eq. (\ref{eq:Photon_path_representation})
we can exploit the fact that for the application of the inverse Laplace
transform we can choose an axis of integration with $\text{Im}\left[\Lambda\right]$
arbitrary large. By doing so, we can make $T_{1}^{a,b}/(\Lambda+\iE{\Gamma}/{2}-\omega_{eg})$
arbitrary small and thereby enforce convergence of the series.
Of course, at first sight it may appear as a major complication to
apply the Neumann series in order to invert a 2 by 2 matrix but due
to the retardation effects in the system caused by the finite speed
of light  we only have to take into account finitely many terms. However,
this expansion enables us to evaluate the inverse Laplace transform
and leads to a semiclassical photon path representation of the relevant
probability amplitudes \cite{Milonni1974,Alber2013}.
Hereby the terms generated by the Neumann expansion in Eq. (\ref{eq:Photon_path_representation}) represent sequences of
spontaneous emission and absorption processes connected by the propagation of single photons.
Thereby, the propagation of a photon from atom $a$ to atom $b$ is described by the function
$T_{1}^{a,b}$.

\subsection{Evaluating the functions $A^{a,b}(\Lambda)$\label{Evaluating_A}}

The evaluation of the functions $A^{a,b}(\Lambda)$ is the main difficulty
in order to analyze the dynamics. Therefore we are going to make use
of the semiclassical results obtained in Sec. \ref{Solving the Helmholtz equation by using semiclassical methods}.
Due to our semiclassical quantization functions we are able to label
the mode functions by the integers $n_{1},\; n_{2}$. Therefore we
obtain 
\begin{equation}
A^{a,b}(\Lambda)=\iE\frac{|D|^{2}}{2\epsilon_{0}\hbar}\sum\limits _{n_{1},n_{2}}\omega_{n_{1},n_{2}}\frac{\left(\mathbf{g}_{n_{1},n_{2}}^{b}\right)_{z}\left(\mathbf{g}_{n_{1},n_{2}}^{a}\right)_{z}^{*}}{\Lambda-\omega_{n_{1},n_{2}}}\;.
\end{equation}
The semiclassical treatment of the mode functions delivers not only
the mode functions for discrete values of $n_{1},n_{2}$ but also smooth
interpolations for all real numbers in between. Of course these
values are unphysical, because they correspond to a violation of the
quantization condition. We can however exploit this smooth interpolation
by applying the Poisson summation formula. Thus, we obtain
\begin{eqnarray}
A^{a,b}(\Lambda)&=&\iE\frac{D^{2}}{2\epsilon_{0}\hbar}\sum\limits _{N_{1},N_{2}\in\mathbb{Z}}\int\limits _{0}^{\infty}\int\limits _{0}^{\infty}\omega(n_{1},n_{2})\frac{\left(\mathbf{g}^{a}(n_{1},n_{2})\right)_{z}^{*}\left(\mathbf{g}^{b}(n_{1},n_{2})\right)_{z}}{\Lambda-\omega(n_{1},n_{2})}e^{\mathrm{i}2\pi(n_{1}N_{1}+n_{2}N_{2})}dn_{1}dn_{2}\nonumber\\
&\approx&\iE\frac{D^{2}}{2\epsilon_{0}\hbar}\sum\limits _{N_{1},N_{2}\in\mathbb{Z}}\int\limits _{-\infty}^{\infty}\int\limits _{-\infty}^{\infty}\omega(n_{1},n_{2})\frac{\left(\mathbf{g}^{a}(n_{1},n_{2})\right)_{z}^{*}\left(\mathbf{g}^{b}(n_{1},n_{2})\right)_{z}}{\Lambda-\omega(n_{1},n_{2})}e^{\mathrm{i}2\pi(n_{1}N_{1}+n_{2}N_{2})}dn_{1}dn_{2}\nonumber\\
&\approx&\iE\frac{D^{2}c_0}{d \epsilon_{0}\hbar}\sum\limits _{N_{1},N_{2}\in\mathbb{N}_{0}}\int\limits _{-\infty}^{\infty}\int\limits _{-\infty}^{\infty}\left|\mathcal{J}\left(\begin{array}{c}
n_{1}(\kappa,\alpha_{n})\\
n_{2}(\kappa,\alpha_{n})
\end{array}\right)\right|\kappa_{eg}\frac{\left(\mathbf{g}^{a}(\kappa_{eg},\alpha_{n})\right)_{z}^{*}\left(\mathbf{g}^{b}(\kappa_{eg},\alpha_{n})\right)_{z}}{\Lambda-2c_{0}\kappa/d}\nonumber\\
&&\cdot e^{\iE2\pi(n_{1}(\kappa,\alpha_{n})N_{1}+n_{2}(\kappa,\alpha_{n})N_{2})}d\kappa d\alpha_{n}\;\nonumber\\
\label{eq:Integral_A}
\end{eqnarray}
with $\mathcal{J}$ denoting the Jacobian determinant. In the second and third step we have used the fact that the
dynamics of the system is mainly influenced by the mode functions
with $\kappa\approx\kappa_{eg}$ and $\alpha_{n}\approx0$. Furthermore we have restricted the sum to nonnegative integers in the third step. 
A numerical calculation confirms that these terms quickly go to zero for increasing values of $d/\lambda_{eg}$ and $f/\lambda_{eg}$ and are already negligible for $d/\lambda_{eg},f/\lambda_{eg}>1$.
In fact these terms are artefacts of the semiclassical approximation.
Finally, we obtain 
\begin{eqnarray}
&A^{1,1}(\Lambda)=A^{2,2}(\Lambda)\nonumber\\
&={\Gamma}/{2}+\sum\limits _{(N_{1},N_{2})\in\mathbb{N}_{0}\times\mathbb{N}_{0}/(0,0)}e^{\iE\tau(N_{1},N_{2})\Lambda}A_{N_{1},N_{2}}\;,\label{eq:A11}\\
&A^{1,2}(\Lambda)=A^{2,1}(\Lambda)\nonumber\\
&=\sum\limits _{N_{1}\in\left[\mathbb{N}_{0}+\frac{1}{2}\right]}\sum\limits _{N_{2}\in\mathbb{N}_{0}}e^{\iE\tau(N_{1},N_{2})\Lambda}A_{N_{1},N_{2}}\;,\label{eq:A12}
\end{eqnarray}
with
\begin{eqnarray}
&A_{N_{1},N_{2}}=\Gamma(-1)^{N_{2}}\mathcal{W}(s_{\alpha_{n}}(N_{1},N_{2}))\;,\label{eq:Gewichtungsfaktoren}\\
\end{eqnarray}
\begin{eqnarray}
&s_{\alpha_{n}}(N_{1},N_{2})=N_{1}\frac{4I_{\mathcal{G}}^{0}}{\sqrt{1-\frac{1}{4\kappa_{eg}^{2}}}}-4N_{2}I_{\mathcal{F}}^{0}\;,\\
&\tau(N_{1},N_{2})=N_{1}{2d}/{c_{0}}+N_{2}{2f}/{c_{0}}\;,\label{eq:tau(N1,N2)}\\
&\mathcal{W}(x)=3\left(-\text{csch}\left(x\right){}^{2}+ x\text{coth}\left( x\right)\text{csch}\left( x\right)^{2}\right)\label{eq:G(x)}
\end{eqnarray}
and with
\begin{eqnarray}
 &\Gamma= \Gamma_{\text{free}} \frac{I_{\mathcal{F}}^{0}+I_{\mathcal{G}}^{0}\frac{f}{d}\sqrt{1-{1}/{(4\kappa_{eg}^{2})}}^{-1}}{(I_{\mathcal{F}}^{2}I_{\mathcal{G}}^{0}-I_{\mathcal{F}}^{0}I_{\mathcal{G}}^{2})}\;.\label{eq:decay_rate}\\
\end{eqnarray}
The spontaneous decay rate in free space is denoted by  
\begin{eqnarray}
  &\Gamma_{\text{free}}=\frac{D^{2}\omega_{eg}^{3}}{3c_{0}^{3}\pi\epsilon_{0}\hbar}\;.
\end{eqnarray}
In fact the constants $A_{N_{1},N_{2}}$ which appear in the functions
$A^{a,b}(\Lambda)$  weight the different photon paths. The corresponding time delays connected to these photon paths caused by the finite speed of light are given by $\tau(N_{1},N_{2})$.
The expression $\Gamma/2$  which appears in Eq. (\ref{eq:A11}) turns out to be the only expression not connected to such a time delay and describes the spontaneous decay of an excited atom.
The spontaneous decay rate $\Gamma$ turns out to deviate from $\Gamma_{\text{free}} $ for $d$ and $f$   below or around $\lambda_{eg}$ but quickly
approaches $\Gamma_{\text{free}} $ as $d/\lambda_{eg}$ and $f/\lambda_{eg}$ increase.
In fact this deviation also turns out to be an artefact introduced by the semiclassical approximations. Thus, we replace $\Gamma$ by $\Gamma_{\text{free}} $ in the following.
Due to the fact that the parabolic cavity is only a special case of a
prolate ellipsoidal cavity we are also able to reproduce the results
of Ref. \cite{Alber2013} by considering the limit $d\rightarrow\infty$,
$f=\text{const.}$ .

\subsection{Photon path representation in the limit of short wavelength $\lambda_{eg}\ll g,f$
\label{short_wavelength}}
In this subsection we are going to investigate the short wavelength limit  $\lambda_{eg}\rightarrow 0$ with
$f=\text{const.}$ and $d=\text{const.}$. In this limit most terms of the
series expansion of the function $A^{a,b}(\Lambda)$ in Eq. (\ref{eq:A11}) and Eq. (\ref{eq:A12})
vanish and
only the expressions with $2N_{1}=N_{2}$
contribute. This directly leads to the delay times $N_2\tau$ with $\tau=(2f+d)/c_{0}$ which we would have expected
by applying the multidimensional JWKB method \cite{Maslov1981}  which is directly connected to the framework of geometrical optics. All the expressions with $2N_{1}\neq N_{2}$ describe diffraction effects which are only relevant
in case of $f$ or $d$ being of the order of a few wavelengths.

Additionally, we observe that in the
mentioned limit the contributing terms are weighted differently. The corresponding  weighting  factor is given by
\begin{eqnarray}
&\left|A_{N_1,N_2}\right|/\Gamma=\delta_{N_{2},2N_{1}}3\frac{N_{2}\log\left(\frac{\epsilon+1}{1-\epsilon}\right)\coth\left(N_{2}\log\left(\frac{\epsilon+1}{1-\epsilon}\right)\right)-1}{\text{sinh}^{2}\left(N_{2}\log\left(\frac{\epsilon+1}{1-\epsilon}\right)\right)}\;\nonumber\\
\label{eq:weighting_function_short_wavelenth}
\end{eqnarray}
with $\epsilon={d}/({d+2f})$ denoting the eccentricity of the cavity
which is a measure of the deviation from a sphere ($\epsilon=0$ corresponds
to a sphere and $\epsilon\rightarrow1$ to a parabola).
The coupling between the atoms increases for $\epsilon\rightarrow0$
(which corresponds to an almost spherical symmetric cavity) and decreases
for $\epsilon\rightarrow1$.
This reduction of the coupling efficiency for increasing $\epsilon$
is caused by a confinement of the wave packets, carrying the excitation from one atom to the other, to the symmetry axis of the cavity.
Therefore for $\epsilon$ close to unity  the electric field at the position of the second atom is almost perpendicular to the symmetry axis and
thus it's coupling to the dipole matrix element of the second atom is reduced.

It is indeed
possible to reproduce all the results we have obtained for the short wavelength limit including the weighting of the different photon paths also
by using the multidimensional JWKB method \cite{Maslov1981} and the framework of geometrical optics.

\section{Results\label{Results}}

\begin{figure}
\centering \includegraphics[scale=0.34]{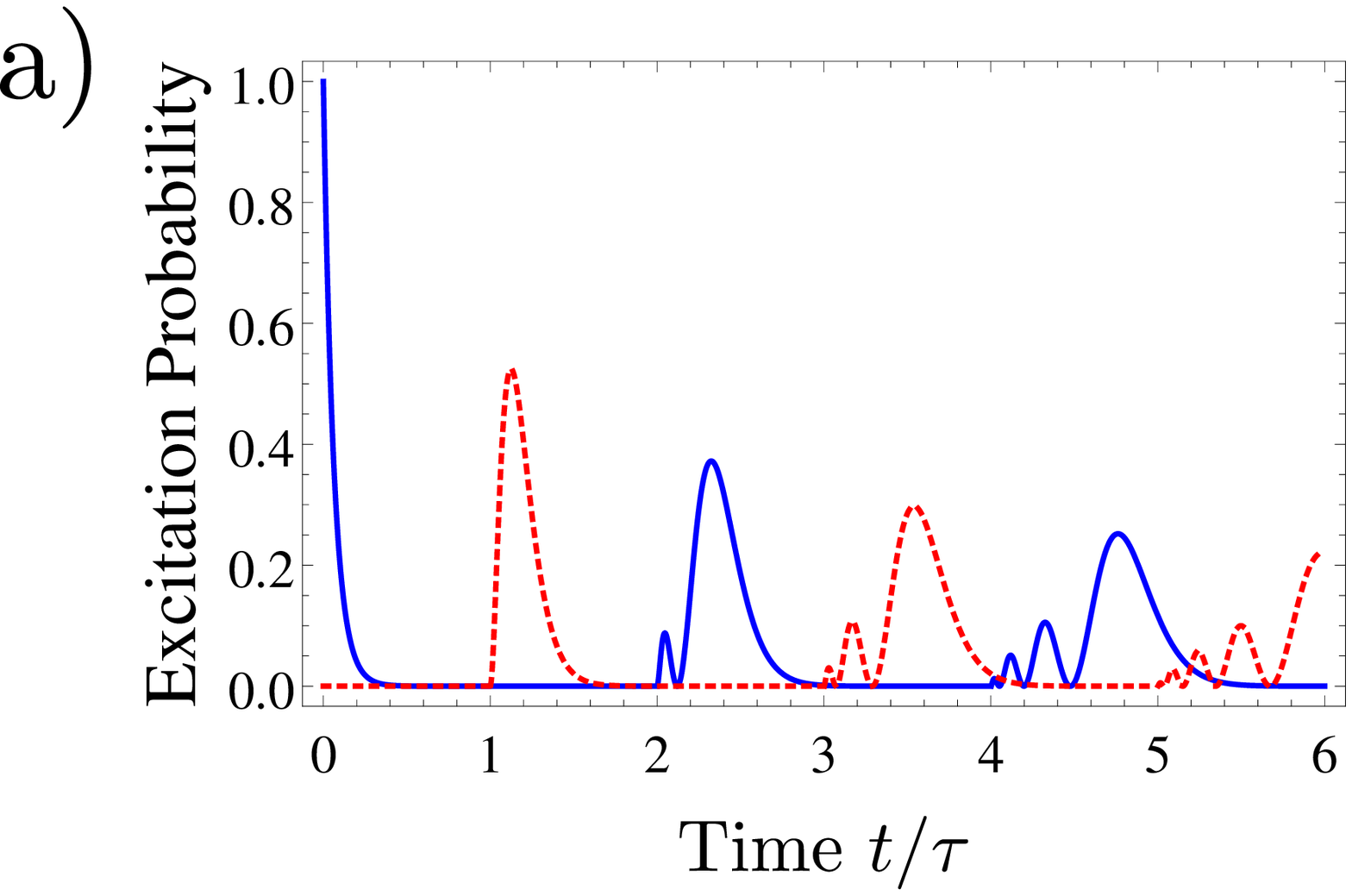} \includegraphics[scale=0.34]{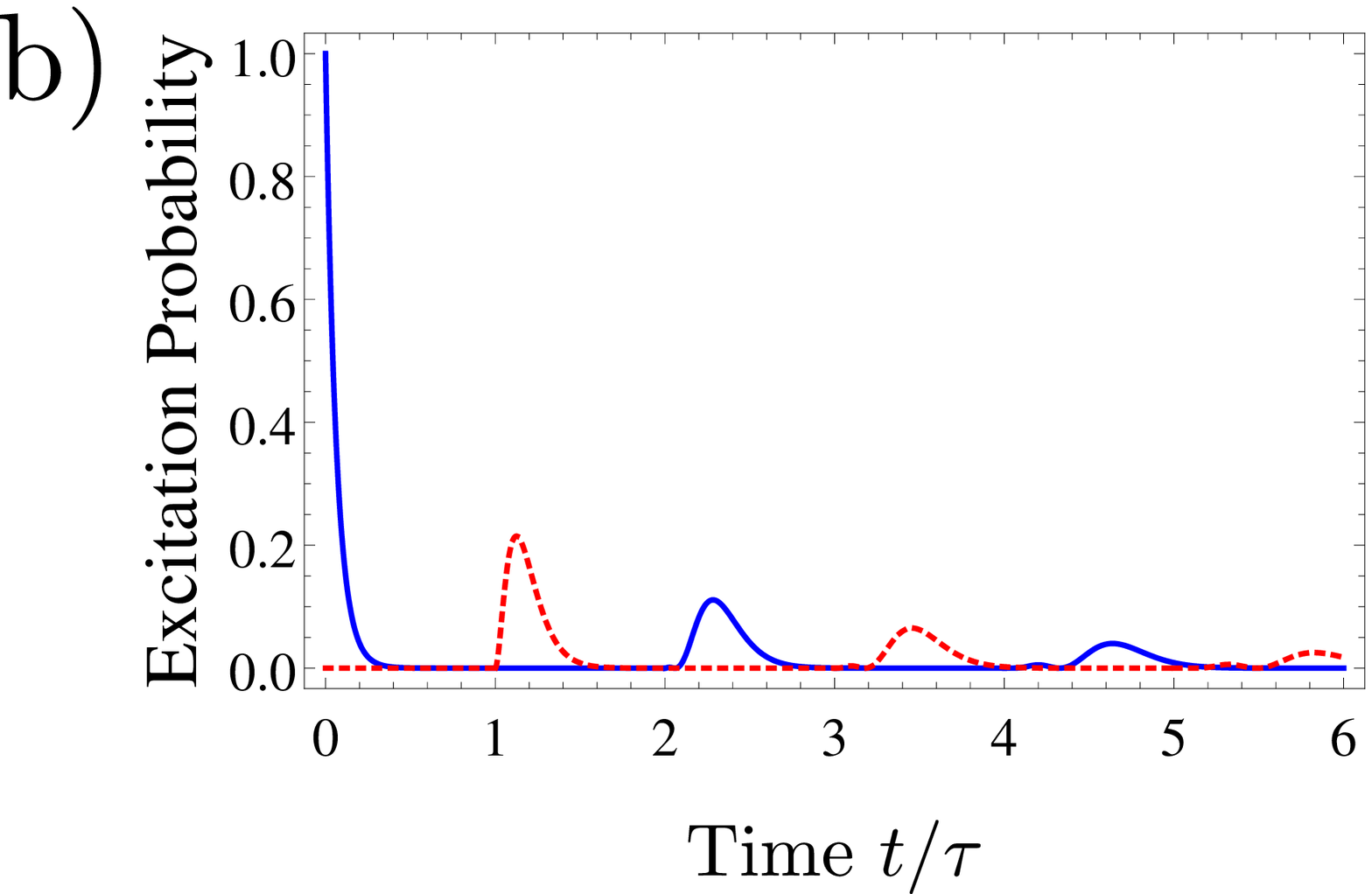}

\includegraphics[scale=0.34]{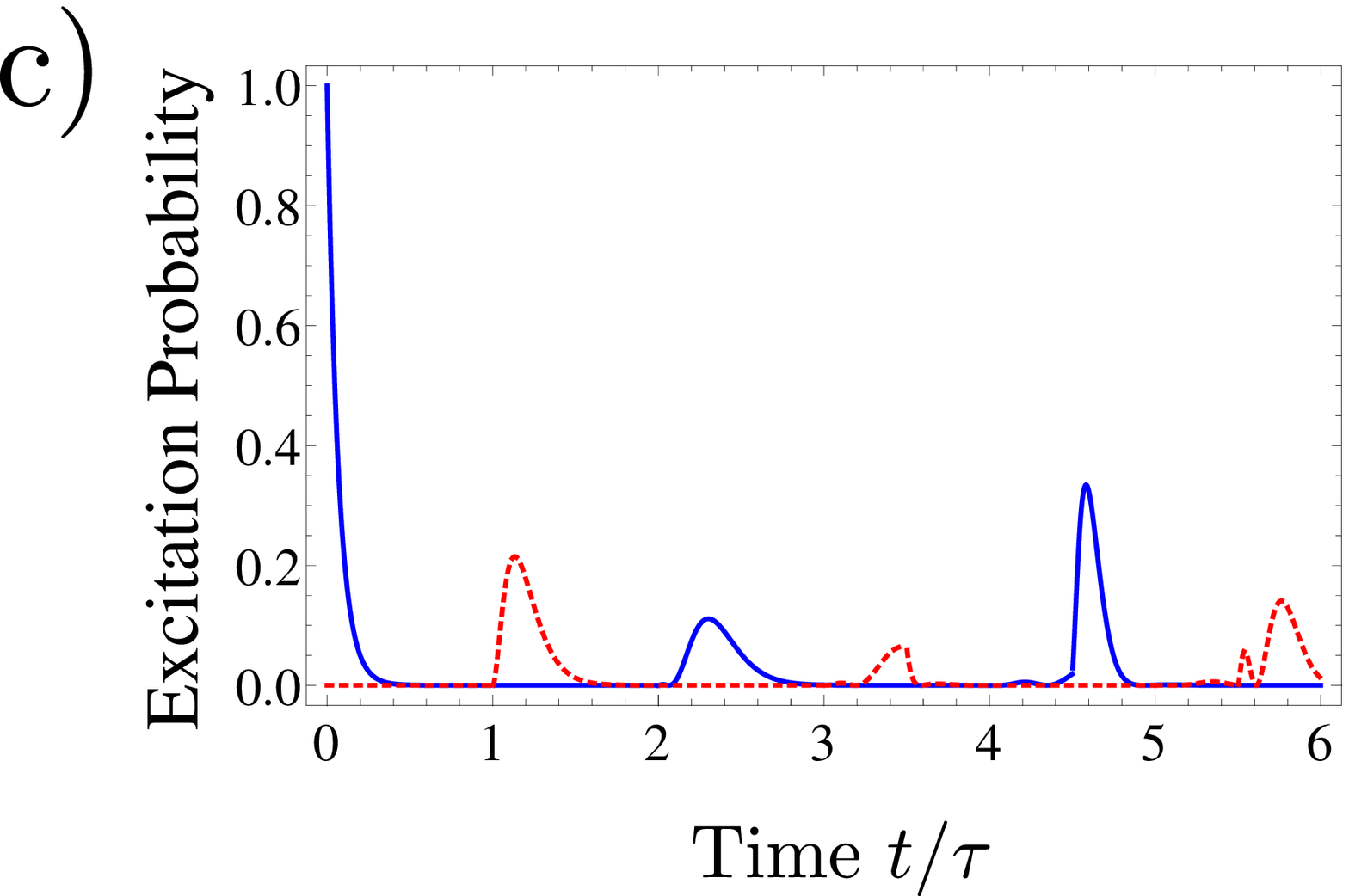}
\caption{Atomic excitation probabilities for $\Gamma\tau=16$
and $\epsilon=1/10,\;\lambda_{eg}\ll d,f$ (a) and $\epsilon=1/2,\;\lambda_{eg}\ll d,f$ (b),  $d/\lambda_{eg}=20,\;f/\lambda_{eg}=10$ (c).
The excitation probability
of atom 1 (2) is represented by the blue (red) curve. \label{fig:Wahrscheinlichkeit_Atomare_Anregung_endliche_Cavity_4}}
\end{figure}

We start our discussion of the dynamics of the atoms by investigating
the extreme multimode scenario which corresponds to $\tau\Gamma\gg1$.
In this scenario the spontaneous decay rate $\Gamma$ is large compared
to the frequency distance between two neighboring modes. Thus the atoms
couple to many modes simultaneously. This leads to a situation which
substantially deviates from the situation in the single mode regime.
The typical dynamics is illustrated in Fig. \ref{fig:Wahrscheinlichkeit_Atomare_Anregung_endliche_Cavity_4}
a and Fig. \ref{fig:Wahrscheinlichkeit_Atomare_Anregung_endliche_Cavity_4}
b. In these two examples, we chose situations with relatively small
wavelength $\lambda_{eg}\ll d,f$ in which the multidimensional JWKB-method
leads to a good description of the system. 

In these cases the excitation probabilities of the atoms are
sharply peaked in time. After the decay of the first atom the second
atom is excited after time $\tau$, which is the typical time
the photon needs to travel from the first atom to the second one as
expected from geometrical optics. The remaining peaks in Fig. \ref{fig:Wahrscheinlichkeit_Atomare_Anregung_endliche_Cavity_4}
a and Fig. \ref{fig:Wahrscheinlichkeit_Atomare_Anregung_endliche_Cavity_4}
b can be understood in terms of descriptive photon paths as discussed in Sec. \ref{Photon_path_representation}. By comparing
the dynamics illustrated in Fig. \ref{fig:Wahrscheinlichkeit_Atomare_Anregung_endliche_Cavity_4}
a with the dynamics illustrated in Fig. \ref{fig:Wahrscheinlichkeit_Atomare_Anregung_endliche_Cavity_4}
b we can study the influence of the geometry of the cavity as already
discussed in Subsec. \ref{short_wavelength}. The situation in Fig.
\ref{fig:Wahrscheinlichkeit_Atomare_Anregung_endliche_Cavity_4} a
with ${f}/{d}={9}/{2}\Leftrightarrow\epsilon={1}/{10}$
corresponds to an almost spherical cavity. Therefore, the coupling
between both atoms is relatively large. The situation in Fig.
\ref{fig:Wahrscheinlichkeit_Atomare_Anregung_endliche_Cavity_4} b
with${f}/{d}={1}/{2}\Leftrightarrow\epsilon={1}/{2}$
corresponds to a highly non spherical cavity. Thus, in accordance
with the results obtained in Subsec. \ref{short_wavelength}, the coupling
between the atoms is reduced.

The semiclassical expressions obtained by the separation
ansatz in Sec. \ref{Solving the Helmholtz equation by using semiclassical methods} can also be used to study
diffraction effects in cases of relatively long wavelengths
($\lambda_{eg}$ of similar order of magnitude as $d$ or $f$).
Such a situation is illustrated in Fig.
\ref{fig:Wahrscheinlichkeit_Atomare_Anregung_endliche_Cavity_4} c.
The influence of diffraction effects can be illustrated by comparing Fig. \ref{fig:Wahrscheinlichkeit_Atomare_Anregung_endliche_Cavity_4} c
and Fig. \ref{fig:Wahrscheinlichkeit_Atomare_Anregung_endliche_Cavity_4} b because both figures correspond to a cavity with eccentricity $\epsilon=1/2$. Due
to these diffraction effects, the dynamics of the system in the regime
$\lambda_{eg}\approx f,g$ is much more complicated than in the regime
$\lambda_{eg}\ll f,g$. In the photon path representation those diffraction
effects are connected with the appearance of additional photon paths
which are not connected to geometrical optics.

\begin{figure}
\centering \includegraphics[scale=0.31]{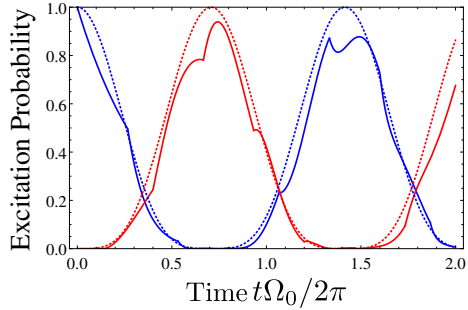}
\caption{Comparison of atomic excitation probabilities for finite $\tau$ with
ideal single mode scenario $\tau\rightarrow0$: The dashed lines correspond to $\tau\rightarrow0$ and the solid lines to the results for finite $\tau$.
The excitation probability
of atom 1 (2) is represented by the blue (red) curves.
The parameters for the solid lines are $\tau=4\pi/15\Omega_0$.
\label{fig:Rabi30}}
\end{figure}

Our former treatment of the extreme multimode scenario also enables us to study the transition from the extreme multimode
scenario which corresponds to $\tau\Gamma\gg1$ to the single mode regime with $\tau\Gamma\ll1$.
To keep the discussion as simple as possible, we
consider the scenario of very small wavelengths and an almost
spherical cavity $\lambda_{eg}\ll d\ll f$ . If we take the limit $\tau\Gamma\rightarrow 0$ with  $\Gamma=\tau\Omega_{0}^{2}/2$ and $e^{\iE2\tau\omega_{eg}}=1$,
we obtain exactly the results of a single mode coupling on resonance to the two two level atoms and $\Omega_{0}$ being the vacuum Rabi frequency of a single atom coupling to this mode.
In Fig. \ref{fig:Rabi30} we compare the single mode scenario for $\tau\rightarrow 0$  with the results for finite $\tau$ which are more complicated due to the presence of additional modes.

\section{Conclusion \label{V. Conclusion}}
We have investigated the dynamics of spontaneous photon emission and absorption processes of
two two-level atoms trapped close to the focal points of a prolate-ellipsoidal cavity.
Our theoretical approach is based on a full multimode treatment of
the electromagnetic radiation field and incorporates the dipole approximation
and the RWA. In order to deal with the electromagnetic radiation field
in the multimode scenario, we have applied semiclassical methods by exploiting
the separability of the relevant electromagnetic mode functions.
With the help of the expressions obtained by our semiclassical treatment,
we have developed a semiclassical photon path representation of the relevant
probability amplitudes. This photon path representation enables us
to discuss the dynamics inside the cavity by means of descriptive
photon paths. This way, we have studied the interplay between both atoms mediated by the radiation
field and have addressed intermediate instances between well-known quantum-optical scenarios. These well-known scenarios include Rabi oscillations which emerge  for $\Gamma\tau\ll1$
or  an almost Markovian dynamics of the atoms dominated
by spontaneous decay processes which emerge as $\Gamma\tau\gg1$ and as the eccentricity $\epsilon$ approaches unity. We have also investigated diffraction effects
and their suppression in the limit of extremely small wavelengths.
\ack
This work is supported by CASED III, the BMBF Project Q.com, and by the DFG as part of the CRC 1119 CROSSING.
\section*{References}
\bibliographystyle{iopart-num}
\bibliography{Literatur.bib}
\end{document}